\documentclass[aps,twocolumn,showpacs,superscriptaddress,floatfix,letter]{revtex4}
\usepackage{graphicx}
\usepackage{psfrag}
\usepackage{color}
\begin{document}
\newcommand{\beq}{\begin{equation}}
\newcommand{\eeq}{\end{equation}}
\newcommand{\ben}{\begin{eqnarray}}
\newcommand{\een}{\end{eqnarray}}
\newcommand{\bea}{\begin{array}}
\newcommand{\eea}{\end{array}}
\newcommand{\om}{(\omega )}
\newcommand{\bef}{\begin{figure}}
\newcommand{\eef}{\end{figure}}
\newcommand{\leg}[1]{\caption{\protect\rm{\protect\footnotesize{#1}}}}
\newcommand{\ew}[1]{\langle{#1}\rangle}
\newcommand{\be}[1]{\mid\!{#1}\!\mid}
\newcommand{\no}{\nonumber}
\newcommand{\etal}{{\em et~al }}
\newcommand{\geff}{g_{\mbox{\it{\scriptsize{eff}}}}}
\newcommand{\da}[1]{{#1}^\dagger}
\newcommand{\cf}{{\it cf.\/}\ }
\newcommand{\ie}{{\it i.e.\/}\ }   
\setlength\abovedisplayskip{5pt}
\setlength\belowdisplayskip{5pt}
\title{Superradiance Transition in Photosynthetic Light-Harvesting Complexes}

\author{G.L.~Celardo$^1$, F.~Borgonovi}
\affiliation{Dipartimento di Matematica e
Fisica and Interdisciplinary Laboratories for Advanced Materials Physics,
 Universit\`a Cattolica, via Musei 41, 25121 Brescia, Italy
and\\
 Istituto Nazionale di Fisica Nucleare,  Sezione di Pavia, 
via Bassi 6, I-27100,  Pavia, Italy}
\author{M. Merkli}
\affiliation{Department of Mathematics and Statistics,  Memorial
University of Newfoundland, St. John's, Newfoundland, Canada A1C
5S7.}
\author{V.I. Tsifrinovich}
\affiliation{Department of Applied Physics, Polytechnic Institute of NYU, 6 MetroTech Center, Brooklyn, NY 11201, USA
}
\author{G.P. Berman}
\affiliation{Theoretical Division, MS B213, Los Alamos National
Laboratory, Los Alamos, NM 87545, USA}

\begin{abstract}                
We investigate the role of long-lasting quantum coherence in the efficiency of energy transport at room temperature
in  Fenna-Matthews-Olson
photosynthetic complexes.
The excitation energy transfer  due to the coupling of the light harvesting 
complex to the reaction 
center  (``sink'') is analyzed using  an effective non-Hermitian
Hamiltonian.  
We show that, as the coupling to the reaction center is varied,
 maximal efficiency in energy transport is achieved
in the vicinity of the  superradiance transition, characterized by a segregation
of the imaginary parts of the eigenvalues of the  
effective non-Hermitian Hamiltonian.
Our results demonstrate that the presence of the sink (which provides 
a quasi--continuum in the energy spectrum) is the dominant effect in
the energy transfer which takes place even in absence of a thermal bath.
This approach  allows one 
to study the effects of finite temperature and the effects of
any coupling scheme to the reaction center. 
Moreover, taking into account a realistic electric dipole interaction,
we show that the optimal distance from the reaction center to the 
Fenna-Matthews-Olson system  occurs at 
the superradiance transition, and we show that this is  consistent with 
available  experimental data.   

\end{abstract}                                                               
                                                                            
\date{\today}
\maketitle

\section{ Introduction} The annual amount of energy humans currently use is
delivered to Earth by the Sun in a few hours! 
Since solar energy is very dilute, it is  essential to transport the captured energy efficiently.
Most natural photosynthetic systems 
consist of antenna complexes, which capture photons from the Sun
and transport energy to a 
reaction center (RC). There,  it is transformed into chemical energy
via charge separation. 
Antenna complexes are able to transfer 
excitations to RCs with an efficiency exceeding $95 \% $. 
For a long time, it was thought that energy transfer in photosynthetic light-harvesting complexes 
occurs through classical processes, similar to random walks of the exciton
to the RC. 
However, surprising evidence of coherent 
quantum energy transfer has been found recently \cite{photo,photoT}.  
These findings raise two basic questions. How can coherence be maintained
in complex biological systems at room  temperature?
Why is quantum coherence relevant to the efficiency of
energy transfer?\\
The first question has been addressed in \cite{photo2,photo3}. We consider here the second one. It is known that quantum coherence can speed up energy transport through a quantum walk, which can be 
faster than a classical walk \cite{qwalk}. 
Although  the relevance of a mechanism 
similar to Dicke superradiance \cite{dicke54}
 has been also pointed out in \cite{srlloyd,sr2}, 
we focus here
on a different feature of the ``superradiance transition'' 
(ST) \cite{SZNPA89,Zannals}. 
We show that ST 
is a dominant mechanism in an antenna complex described by discrete
energy levels coupled to the RC,  modeled here by a sink having a continuum
energy spectrum similar to what has been done in \cite{photo2,qwalk,cechi}.
On the other hand, the effects of the thermal bath lead only to small
corrections to the energy transport in the vicinity of maximal efficiency.
The antenna-sink
coupling 
causes the appearance of a resonance
width (inverse of life-time) and 
an energy (Lamb) shift.
For weak coupling strength, 
the resonance widths  are roughly the same. However, 
if the coupling strength reaches a critical value, at which the resonance widths start to overlap, then a segregation of widths builds up. In this regime, almost the entire (summed up) decay width is allocated to just a few short-lived ``superradiant states'', while all other states are long-lived (and effectively decoupled from the environment).
We call this segregation the  ``Superradiance Transition''. This effect
 has been studied using random matrix theory 
\cite{verbaarschot85,puebla}, in nuclear physics \cite{Volya}, 
for microwave billiards \cite{rotter2} and in paradigmatic models 
of coherent quantum transport \cite{kaplan,rottertb}.
It was shown in \cite{kaplan} that in a realistic model for quantum transport,  maximum transmission is achieved at ST.\\
In this paper, we focus on transport properties of the Fenna-Matthews-Olson (FMO)
complex, found in green sulphur bacteria.
This complex, one of the most 
studied in the literature \cite{ photo2,photo3,qwalk,renaud}, acts as a conductor  for energy transport 
between the antenna system and the RC. The FMO complex is a dissipative 
open quantum system
which interacts with the thermal bath provided by the protein environment.
Here we take an effective non-Hermitian 
Hamiltonian approach \cite{MW,SZNPA89,Zannals} and study the ST 
as a function of the coupling to the RC
and the thermal bath, due to phonons.
The phonon bath induces  dephasing and dissipation,
and we take both effects into account using two different models for the phonon bath. 
Since ST is due to quantum coherence,  
we address here two main issues: i)  whether its effects can survive in presence of
 dephasing induced by the phonon bath
 at room temperature, and ii) how ST depends on the strength of the coupling
between the FMO and the RC.
It has been shown recently that maximal transport efficiency for the 
FMO complex is achieved near a critical  coupling to the 
RC \cite{mohseni}. 
However, so far,  
the dependence of this critical coupling on the parameters of the FMO
and the RC
has not been determined. 
We compute this critical coupling 
 analytically and  show that it 
corresponds to the ST. 
We demonstrate that the quantum coherent effect of ST, 
even taking into account  dephasing and relaxation,
determines the maximal transport efficiency at room temperature.
Indeed, the ST is 
due to coherent constructive interference between the various paths to the RC, 
thus enhancing  the rate of energy transfer.
Finally, with the aid of the non-Hermitian Hamiltonian
approach, we consider a realistic coupling between the FMO complex and the RC,
showing that the ST determines the optimal distance from the RC to the FMO system.   \\

\section{The Model for Superradiance transition }
The FMO complex is a trimer, composed of identical subunits, each of which 
contains seven bacteriochlorophylls (BChl)  \cite{8Bchl}. Each subunit acts 
independently and  can be modelled using a tight-binding Hamiltonian,
\begin{equation}
H_0= \sum_{i=1}^7 E_i |i\rangle \langle i| + 
\sum_{i,j}( J_{i,j} |i \rangle \langle j| + h.c.).
\label{ham0}
\end{equation}
Here, $|i \rangle$ is the state in which 
the 
$i$-th  site is excited and the  others are in the ground state.
  Since the solar energy is very dilute,
we limit the description to a single excitation in the complex, 
as is commonly done in the literature. 
The numerical values of $E_i$ and $J_{i,j}$  have
been taken from Ref.~\cite{photo2}.
Below we take
the matrix elements  of $H_0$ expressed in $cm^{-1}$:

$$
\left( \matrix{ 
200  & -87.7 & 5.5  & -5.9 & 6.7 & -13.7 & -9.9  \cr 
-87.7  & 320 & 30.8  & 8.2 & 0.7 & 11.8 & 4.3  \cr
5.5  & 30.8 & 0  & -53.5 & -2.2 & -9.6 & 6  \cr
-5.9  & 8.2 & -53.5  & 110 & -70.7 & -17 & -63.3  \cr
6.7  & 0.7 & -2.2  & -70.7 & 270 & 81.1 & -1.3  \cr
-13.7  & 11.8 & -9.6  & -17 & 81.1 & 420 & 39.7 \cr
-9.9  & 4.3 & 6  & -63.3 & -1.3 & 39.7 & 230  \cr
} \right) 
$$
The incident photon creates an electron-hole pair, called an exciton, 
which decays due to two processes: coupling to the electromagnetic field, i.e.
emission of a photon (recombination) with an associated decay time, $T_1$, 
and coupling to the RC with a decay  time, $T_{1r}$.\\
As is common in quantum optics \cite{milonni}, we describe 
this dissipative system with at most one excitation by states 
\begin{equation}
\label{cho}
|\psi \rangle  = \sum_{i=1}^7 a_i |0\rangle \otimes |i\rangle  + 
\sum_c \int dE \ b_c(E) |c,E\rangle  \otimes |gs\rangle,
\end{equation}
where $|0\rangle$
 is the vacuum state of the environment and
$|c,E\rangle  \otimes |gs \rangle $ is the state with one excitation 
in the environment and none on the sites. 
Here, $c$, 
 is the quantum number labelling channels (at energies $E$) in the environments.
The reduced density matrix is obtained by  tracing over the states $|0\rangle$ and $|c,E \rangle $,
\begin{eqnarray}
\rho
&=&\sum_{i,j} a_ia_j^* |i \rangle \langle j| + (1- \sum_i |a_i|^2) 
|gs \rangle \langle gs| ,
\label{ro1}
\end{eqnarray}
which  is an $8\times 8$ matrix.
However, $\langle gs|\rho|i\rangle = 0$ since with
the choice  (\ref{cho}), we neglect the 
transitions $|i\rangle \to |gs\rangle$. 
Moreover, $\langle gs|\rho|gs\rangle$ is simply  the loss of probability of excitation 
of the seven sites. Therefore, we restrict our considerations to the 
$7\times 7$ matrix $\langle i |\rho |j\rangle$, $1\leq i,j\leq 7$, 
which however does not have constant trace. \\
In order to compute the evolution of the reduced density
matrix, we introduce 
an effective non-Hermitian Hamiltonian
 \cite{SZNPA89,kaplan,messiah}
which in general 
can be written 
as,  $H_{\mathrm{eff}} (E) = H_0 + \Delta (E) -i W(E)$, 
where $H_0$ is the Hermitian Hamiltonian of the system
decoupled from the environments and 
$\Delta (E)$ and $W(E)$ are the induced energy shift and the dissipation, respectively.
Neglecting the energy dependence and the energy shift
we have 
\begin{equation}
H_{\mathrm{eff}}=H_0 -i W  \hspace{0.5cm} {\rm with}
 \hspace{0.5cm} W_{ij} = \sum_c A_i^c (A_j^c)^*.
\label{Heff}
\end{equation} 
The real symmetric matrix,  $W$, is given in terms
of the bound state-continuum transition amplitudes,
$A_i^c$, from the discrete state $i$ to the continuum channel $c$.\\ 
The Schr\"odinger equation and Eq.~(\ref{cho}), result in the
following equation for the coefficients,
$a_j$:
\begin{equation}
i \hbar \dot{a}_j = \sum_{k=1}^7 \left(
{H_0}_{jk}  a_k -i  W_{jk} a_k\right),
\label{cc}
\end{equation}
and from this the master equation easily follows,
\begin{equation}   
i   \hbar \dot{\rho}_{jk} =  [
H_0\, , \rho ]_{jk} - i \sum_{l=1}^7 \left(W_{jl}
\rho_{lk} + \rho_{jl} W_{lk}\right).  
\label{ro2}
\end{equation}
Under the standard assumption \cite{lloyd,deph} that
each site is coupled to an independent (local)
environment, with associated coupling time $T_1$, we have 
$A_i^i=\sqrt{\hbar/2T_1}$, $i=1,..,7$. The site $i=3$ is the only 
one which is, in addition, coupled to the RC, giving rise to a 
decay time $T_{1r}$. Then $A_3^{8}=\sqrt{\hbar/2T_{1r}}$ 
(in this scheme there are 7+1 channels);
for  other channels  we have $A_i^c=0$.
In Eq.~(\ref{ro2}) we take into account: (i) the interaction 
between the FMO and the RC through the time $T_{1r}$ and
(ii) the characteristic time of exciton recombination,
 $T_1$. The effects of the thermal bath will be considered in Sec.(\ref{sec4}).

One can verify that equation (\ref{ro2}) can also be
 obtained by restriction to the $7\times 7$
 density matrix, from a Lindblad dynamics 
for the full $8\times 8$ density matrix (\ref{ro1}).\\ 
In the following, we fix $T_1=1 \ ns$, which is the exciton recombination time
reported in the literature \cite{lloyd,deph}, and we focus on the effect of
varying $T_{1r}$. \\

\section{ Superradiance transition }
\begin{figure}[ht!]
\vspace{0cm}
\includegraphics[width=8cm,angle=0]{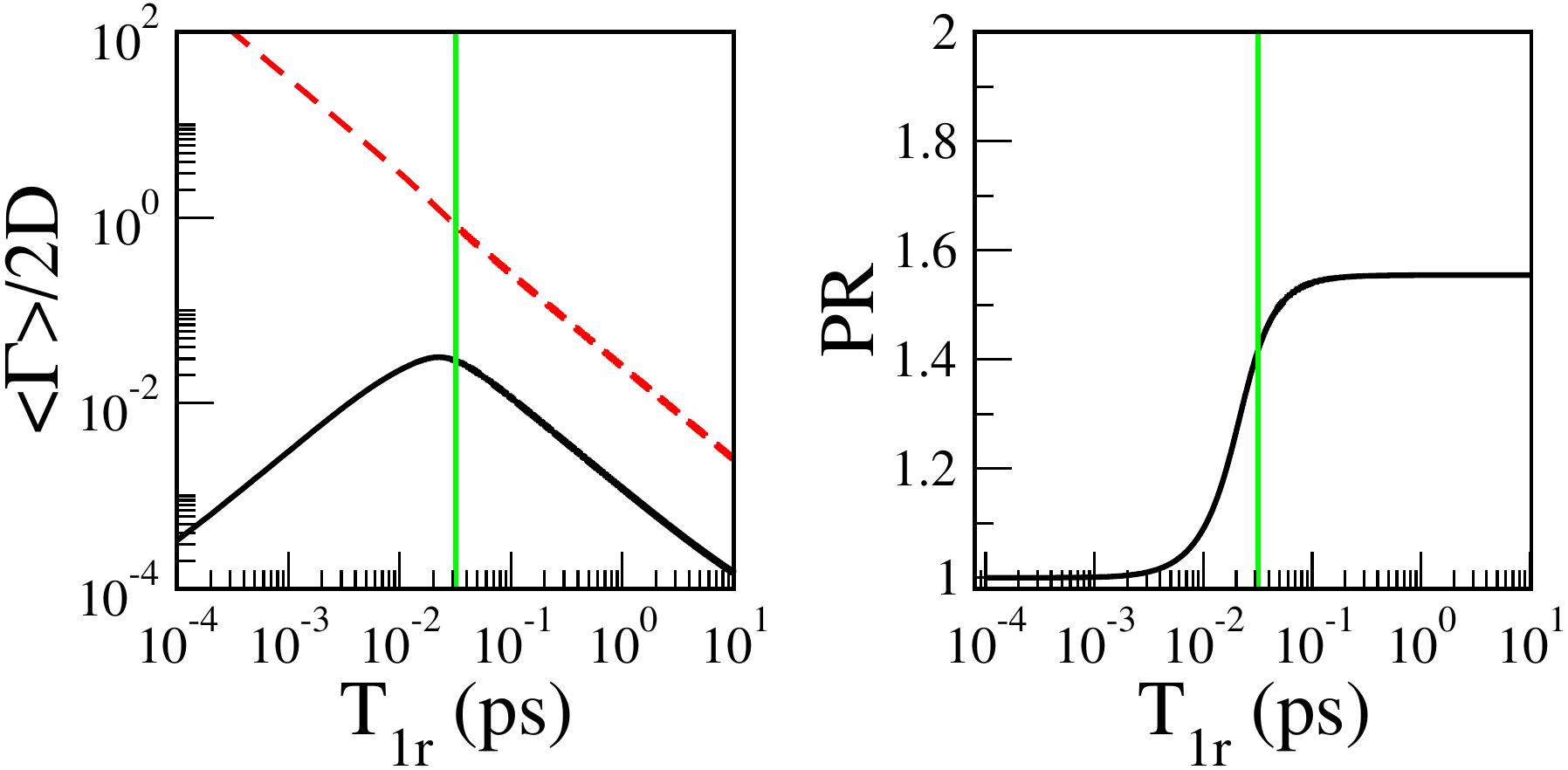}
\caption{ (Color online)
Left panel: average decay width, normalized to the 
mean level spacing, $D$
as a function of the coupling time to the RC,  $T_{1r}$. 
The black curve represents
the average decay width of the $6$ states with smallest width, 
while the dashed red curve shows the
largest decay width. 
Right panel: PR  of the eigenstate
of  $H_{{\mathrm{eff}}}$  with the largest width 
 as a function of  $T_{1r}$. 
In both panels
the vertical (green) line indicates the critical value of $T_{1r}$ 
at which ST
occurs.
}
\label{srloc3}
\end{figure}

ST can be analyzed by studying the complex eigenvalues, 
${\cal E}_r = E_r -i \Gamma_r /2$ of $H_{\rm eff}$,
defined in (\ref{Heff}). 
As the coupling between the excitonic states and the RC increases, 
one observes a rearrangement of the widths, 
$\Gamma_r$ (the  ``superradiance'' transition \cite{kaplan}).
We show this effect in Fig.~\ref{srloc3} (left panel), where 
the largest width  (red dashed curve) and  
the average of the $7-1=6$ smallest 
widths (black full curve) 
are plotted as functions of $T_{1r}$. 
For weak coupling to RC (large $T_{1r}$) the widths of all states
increase as $T_{1r}$ decreases. On the other hand,
below a critical value $T_{1r}^{cr}$, corresponding to ST,
(vertical  line), the average of the $6$ smallest widths 
decreases while 
the largest width, corresponding to the superradiant state,
increases.  To examine localization of the excitation 
we use the  Participation Ratio (PR)\cite{felix}
 of a state $| \psi \rangle$, defined as:
$$
PR=( \sum_i |\langle i|\psi \rangle|^4)^{-1}.
$$
Its value varies from $1$ for fully localized to $7$ for 
fully delocalized states. The right panel of Fig.~\ref{srloc3} 
shows the PR for the state associated with the largest 
width (the one decaying most quickly). In the superradiant 
regime, $T_{1r}<T_{1r}^{cr}$, this state is fully localized on site $3$, the only site connected to the RC. For weak coupling to the RC, $T_{1r}>T_{1r}^{cr}$, the PR is approximately $1.6$. This small value, as compared to the maximal possible value of $7$, is explained by 
(Anderson)  localization \cite{Anderson}  of the eigenstates on sites.  The 
Anderson localization effect in the FMO system is due to the fact
that the excitation energies of the sites and the couplings among them
are all  different. Thus, the FMO complex can be thought of as
a disordered system. 
The critical value, $T_{1r}^{cr}$, at which ST occurs,
 can be estimated analytically.
If all  states have roughly the same width, at least for small
coupling, then the superradiance condition coincides with that of 
overlapping resonances.
Such a reasoning can be applied to the FMO
system, too. Here, eigenstates are mostly localized on the sites, 
and only site $3$ is coupled to the RC. 
The widths are thus not uniform and most of the
total width belongs to the eigenstate localized at site $3$.
Imposing that the half width, $\Gamma_3/2$, is approximately equal
to the mean level spacing $D$, $\Gamma_3/2 \approx D$, and
using   $\Gamma_3 
\approx \hbar/ T_{1r}$, 
we obtain the critical value at which ST occurs,
\begin{equation}
T_{1r}^{cr} \approx \frac{\hbar}{2 D}. 
\label{sr}
\end{equation}
In the FMO system, the energy level spacing  is
$D/hc \approx 83.5$ $cm^{-1}$, which gives $T_{1r}^{cr} \approx 0.03 \ ps$, 
a value in very good agreement with the numerical 
results of Fig.~\ref{srloc3} (vertical  line).\\

Such a value, $T_{1r}=0.03 \ ps$,   corresponds to  a transfer rate, $\kappa$, from site $3$
to the RC of $\kappa = \frac{1}{2 T_{1r}} = 16.6 \ ps^{-1}$.
This value  is 
larger than the values usually mentioned in literature, which 
range from $0.25 \ ps^{-1}$ to $4 \  ps^{-1}$ \cite{deph},
even if  $1 \ ps^{-1}$ is the most common value \cite{qwalk}.  
This discrepancy can be due to the simplicity of our model,
even if it is important to notice that, to the
best of our knowledge,  the real value of
the coupling time, $T_{1r}$, is not exactly known.
In any case, for this reason, in Section VIb, we  consider 
a more realistic coupling scheme between the FMO system and the RC.

\section{Efficiency of energy transport in presence  of a thermal bath} 
\label{sec4}

Interaction with the phonon environment is  complicated, 
and it involves both dephasing and dissipation~\cite{photo2}. 
Since superradiance is due to quantum coherence,
in Sec (\ref{secA})  we first focus on dephasing
and the  consistent indirect relaxation, induced by the presence of classical noise.
On the other hand, in Sec. (\ref{secB}),  we consider both dephasing and dissipation
induced by a finite temperature bath. 
Needless to say, while the latter bath induces at  equilibrium a Gibbs energy
level  distribution, 
the former gives rise to an equal population of all energy levels.

\subsection{Efficiency of energy transport in presence of noise}
\label{secA}

As a first step, one can study the effects of the phonon bath  modelling the
thermal bath by a classical noise. In this case, the dephasing
effects are adequately described using an interaction 
as in Ref.~\cite{lloyd}: 
\begin{equation}
H_{_{SB}}= \sum q_i(t) |i\rangle \langle i| ,
\label{lloyd}
\end{equation}
with 
\begin{equation}
\langle q_i(t) q_j(t) \rangle = \hbar^2 \gamma_d  \delta_{i,j} 
\delta(t),
\label{lloyd1}
\end{equation}
where $\gamma_d$ plays the role of the dephasing rate.
This approach corresponds to an effective infinite
temperature that leads to equal populations of energy levels at sufficiently
large times.
We take into account the interaction (\ref{lloyd})
by adding a dephasing Lindblad operator  to the master equation (\ref{ro2}), 
as was done in Ref.~[19]. 
The interaction with noise  leads to the Haken-Strobl master equation for the 
density matrix of the following form: 
\begin{equation}
\frac{d \rho_{i,j}}{dt}= -\frac{i}{\hbar} \left(
H_{\mathrm{eff}} \rho  - \rho H_{\mathrm{eff}}^{+}
\right)_{i,j}  - \gamma_d(1-\delta_{i,j}) \rho_{i,j}.
\label{HS}
\end{equation}
The first term in the r.h.s. of Eq.~(\ref{HS}) takes into account the coherent 
evolution and the dissipation by recombination and trapping
into  the reaction center, (it is simply Eq.~(\ref{ro2}) rewritten
in terms of  $H_{\mathrm{eff}}$, defined in Eq.~(\ref{Heff})). 
The last term, which corresponds to the decay 
of the off-diagonal matrix elements,
 takes into account dephasing 
and indirect relaxation.
For the FMO system, the amplitude of noise, $\hbar \gamma_d$, is related to the temperature,
$T$, by the relation, found in experiments~\cite{photoT},
\begin{equation}
\hbar \gamma_d (T) \simeq 0.52 \ \hbar c \ (T /K) (cm)^{-1},
\label{gammad}
\end{equation}
where $T$ is the  temperature expressed in kelvin degrees, $K$.

\begin{figure}[ht!]
\vspace{0.cm}
\includegraphics[width=8cm,angle=0]{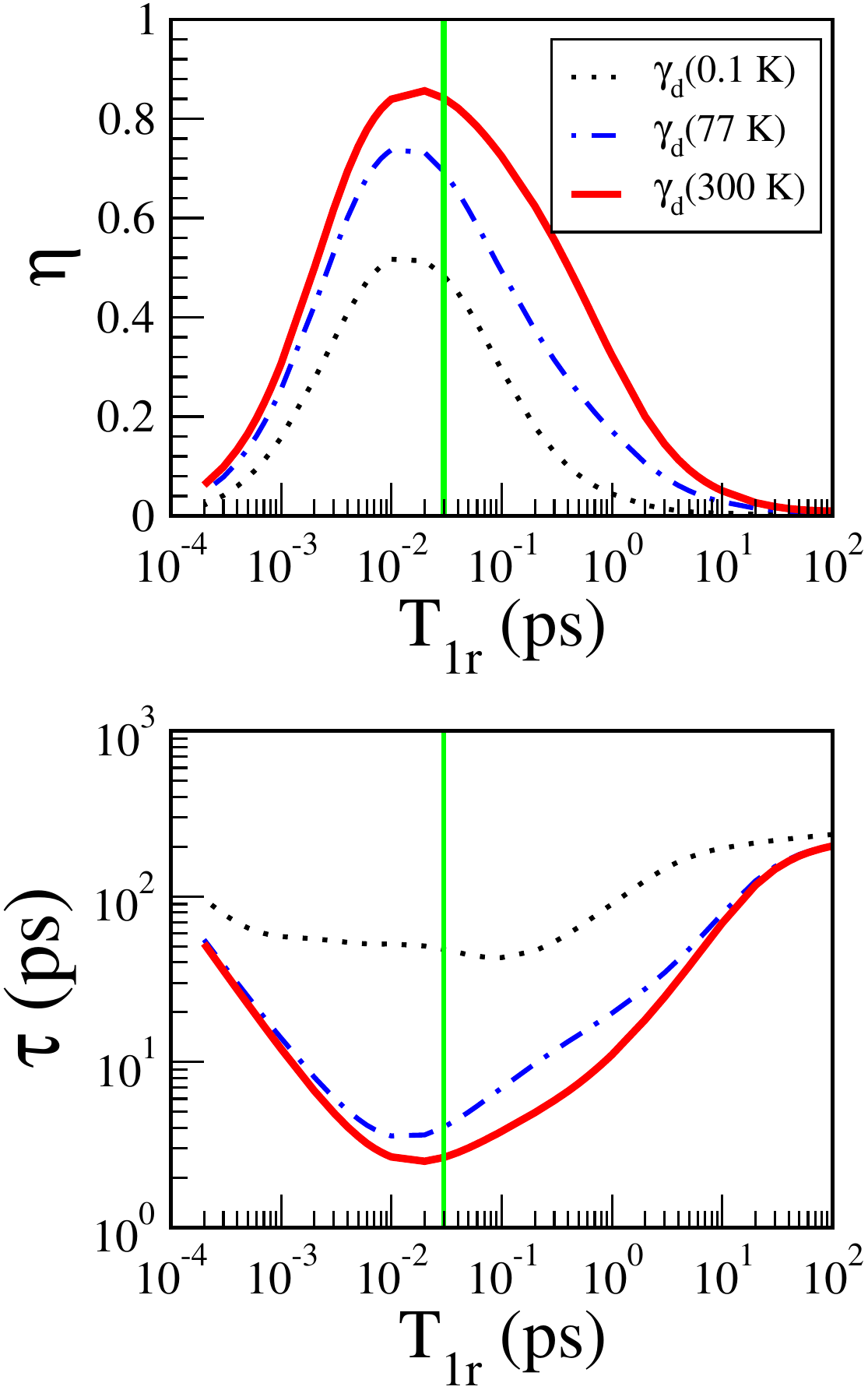}
\caption{ (Color online)
Upper panel : efficiency computed at $t_{max} = 5 \ ps$,
 see Eq.~(\ref{eta}), 
as a function of
 $T_{1r}$, for different dephasing rates, see Eq.~(\ref{gammad}). 
Lower panel:  average transfer time, see Eq.~(\ref{tau}),  as a
 function of $T_{1r}$, for the same effective temperatures.
 ST  has been 
indicated as a green vertical line.
The initial condition is
$\rho(0)= (1/2)(|1 \rangle \langle 1| +|6 \rangle \langle 6|)$.
}
\label{fig2}
\end{figure}

Transport efficiency has been measured
as in Ref.~\cite{deph}
 by the probability that the excitation is in the RC at the time $t_{max}$,
\begin{equation}
\eta(t_{max})= \frac{1}{T_{1r}} \int_0^{t_{max}} \ dt \ \rho_{33}(t),
\label{eta}
\end{equation}
and by the average transfer time to the RC~\cite{lloyd},
\begin{equation}
\tau= \frac{1}{T_{1r}}  \int_0^{\infty} \ dt \, t \,  \rho_{33}(t)/\eta(\infty).
\label{tau}
\end{equation}
In our simulations, we take the initial state
$$
\rho(0)= \frac{1}{2}(|1 \rangle \langle 1| +|6 \rangle \langle 6|),
$$ 
since sites 1 and 6 receive the excitation from the 
antenna system~\cite{lloyd}.\\   
It was numerically found 
 in \cite{mohseni} that the efficiency reaches a maximum as a function of $T_{1r}$. 
Here, we explain this as
a consequence of ST, 
a general phenomenon in coherent quantum transport.
In Fig.~\ref{fig2}, we plot $\eta(t_{max}=5 \, ps)$ (upper panel), 
and $\tau$ (lower panel), as  functions 
of $T_{1r}$. The maximum efficiency of energy transport
 (maximum $\eta$ and minimum $\tau$) is reached near the ST (vertical line). 
Note that $\eta(t_{max}=5 \, ps)$ has a maximum not only in the quantum limit 
($\gamma_d(T=0.1 \, K)$, black dashed curve), but also considering the dephasing rate at 
room temperature ($\gamma_d(T=300 \, K)$, red thick curve). 
These results show that the effects of the 
ST persist even in presence of dephasing and indirect relaxation.
Within the framework of  the ST, the decrease in efficiency for large coupling
to the RC, can be interpreted as a localization effect, see Fig.~\ref{srloc3}
(right panel).
Our results also show that
dephasing can increase efficiency, 
since it counteracts quantum localization.
This effect is known as Environment-Assisted Quantum Transport 
(ENAQT)~\cite{lloyd,deph}.
The average transfer time, see Fig.~\ref{fig2} (lower panel), has a minimum near the ST of the
order of a few picoseconds. This time is comparable
with the  transfer times estimated in the literature.
\vspace{0.8cm}

The coupling to the RC also induces 
a shift of the energy of site $3$ (not only
a decay width) \cite{kaplan}.  
This shift is  assumed to be generically of the 
form $\delta= \epsilon/T_{1r}$, where $\epsilon$ 
depends on the details of the coupling. We checked that the 
effect of changing $\epsilon$ randomly, so as to produce 
up to a  $50 \% $ change in the average level spacing, merely changes 
the efficiency at most by a few percent.\\

\subsection{Efficiency of energy transport in the presence of a finite temperature thermal bath}
\label{secB}

In this subection, we consider the effects 
of energy transport to the RC taking into consideration the interaction
with a phonon bath at finite temperature, $T$, as described in Ref.~\cite{qwalk}.
Here we consider that only site $3$ is coupled to the RC, as described above.
The Lindblad-type master equation has the form 
\begin{equation}
\frac{d \rho_{i,j}}{dt}= -\frac{i}{\hbar} \left(
H_{\mathrm{eff}} \rho  - \rho H_{\mathrm{eff}}^{+}
\right)_{i,j}  + L_p (\rho)_{ij},
\label{lind}
\end{equation}
where the action of the Lindblad operator on $\rho$, $ L_p (\rho)$,
 is described in Eq.~(5) 
of Ref.~\cite{qwalk}. With this choice, at sufficiently large time, the transition to the
Gibbs distribution occurs, in absence of any other dissipative mechanism, such as the presence
of ``sinks''.

\begin{figure}[ht!]
\vspace{0.cm}
\includegraphics[width=8cm,angle=0]{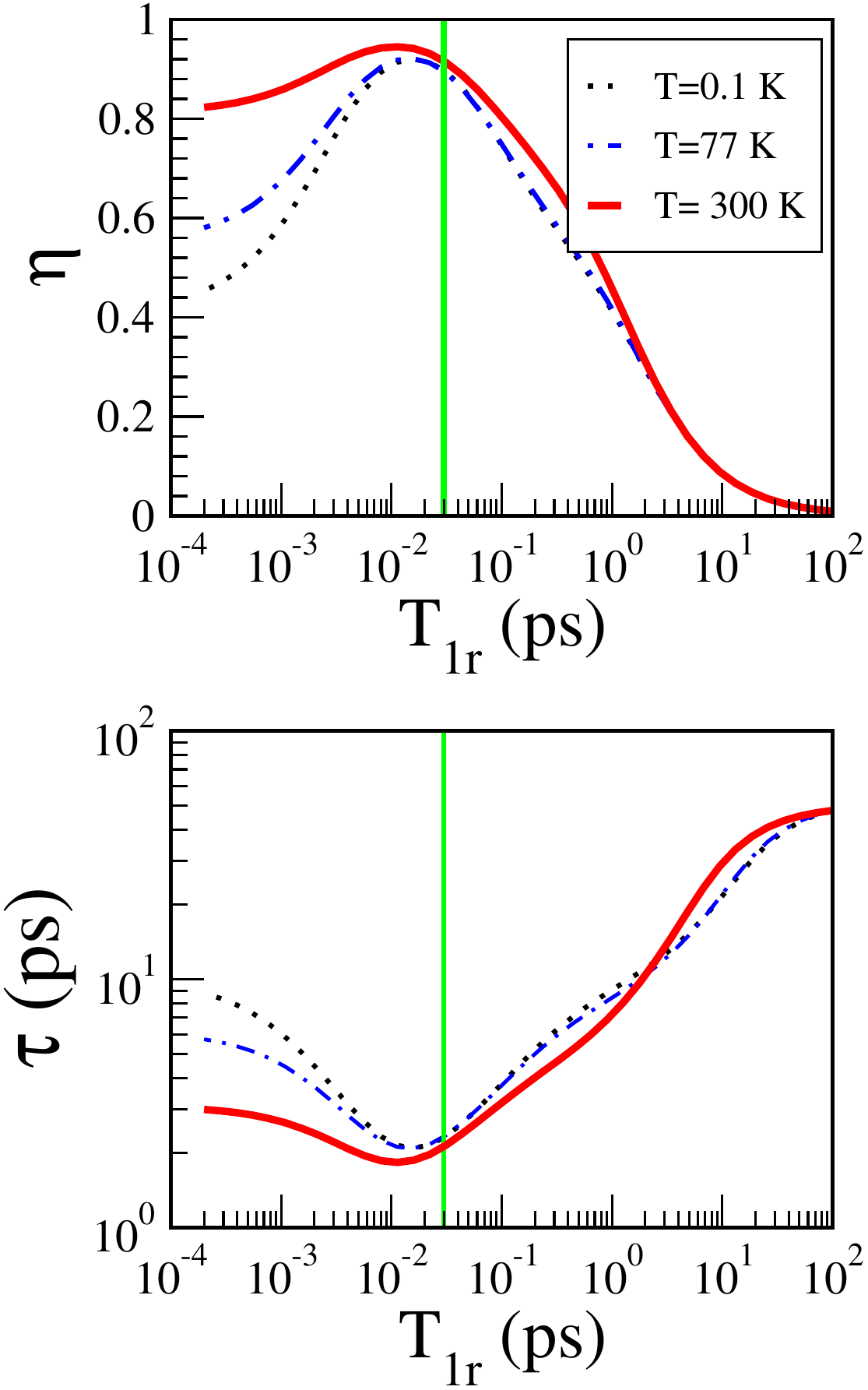}
\caption{ (Color online)
Upper panel : efficiency computed at $t_{max} = 5 \ ps$,
 see Eq.~(\ref{eta}),
as a function of
 $T_{1r}$, for different 
temperatures of the phonon bath.
Lower panel:  average transfer time, see Eq.~(\ref{tau}),  as a
 function of $T_{1r}$, for the same  temperatures.
 ST  has been
indicated as a green vertical line.
The initial condition is
$\rho(0)= (1/2)(|1 \rangle \langle 1| +|6 \rangle \langle 6|)$.
We use as cut-off frequency of the spectral density
of the thermal bath $\omega_c = 150 \ (cm)^{-1}$, and 
a reorganization energy $E_R = 35 (cm)^{-1}$, where the latter
two quantities have been defined in Ref.~\cite{qwalk}.
}
\label{fig2new}
\end{figure}

In Fig.~\ref{fig2new} (upper panel), we present our results on the 
dependence of efficiency, $\eta$, as a function of $T_{1r}$,
for three bath temperatures. 
As one can see,  at  the ST, indicated
as a vertical green line, the efficiency is about $0.92$, and 
it weakly depends on temperature. We also mention that the maximal
efficiency occurs near the  ST.

The value of $T_{1r}$ at which one gets the maximal efficiency should
not be confused with the average transfer time,  see Eq.~(\ref{tau}).
In particular, in Fig.~\ref{fig2new} lower panel, for parameters corresponding
to the ST and room temperature, $T_{1r} \approx 30 \ fs$ and 
$\tau \approx 2.1 \ ps $.

Comparing  Fig.~\ref{fig2} and  \ref{fig2new}, (upper panels), 
one can see that the
presence of the phonon bath 
significantly increases the efficiency
of energy transport to the RC, almost without changing the position 
of its maximum.
We also would like to mention that in the  presence of the thermal bath,
the temperature effects on the efficiency  
are less significant than in presence of a classical 
noise, as in  Fig.~\ref{fig2}. 
 Indeed dissipation helps the system to
reach the site $3$ which has the lowest energy.

The analysis of this Section shows that the consequences of the ST are very
important even in presence of dephasing and dissipation. 
For both models of thermal bath considered in this Section, the ST 
provides the maximal efficiency of energy transport. 
In the following  we will consider only the model of the phonon 
bath presented in Sec.~(\ref{secA}), which, as shown in this Section, is sufficient to capture the main
effects due to  the phonon bath.

\section{ Quantum vs. classical }
ST implies the presence of a maximum of the energy transport
 efficiency as a function of the coupling time
 to the RC, $T_{1r}$. This effect is counter-intuitive
from a classical point of view. Indeed, the probability to escape 
(decay to the RC) for a classical particle does not decrease as 
the escape rate ($1/T_{1r}$ from site $3$) is increased.  
In order to demonstrate the difference between the effects of
quantum coherence on energy transfer discussed above, and 
the corresponding classical energy transport, we consider a 
classical master equation
for the population dynamics, as in the Forster approach \cite{forster}:
\begin{equation}
\frac{dP_i}{dt}= \sum_k  ( T_{i,k} P_k - T_{k,i} P_i )   - 
\frac{P_i}{T_1} - \delta_{i,3} \frac{P_i}{T_{1r}},
\label{master}
\end{equation} 
where $P_i$ is the probability to be on site $i$, $T_{i,k}$
is the transition matrix and the last two terms take into account
the possibility for the classical excitation to escape the system. 
The transition
rates from site $i$ to site $k$ 
have been computed from  \cite{leegwater},
neglecting 
the dependence on the 
coupling to the RC (for a classical particle,
the probability to go from any site to site $3$ does not
depend on the coupling to the RC).
\begin{figure}[ht!]
\vspace{0cm}
\includegraphics[width=8cm,angle=0]{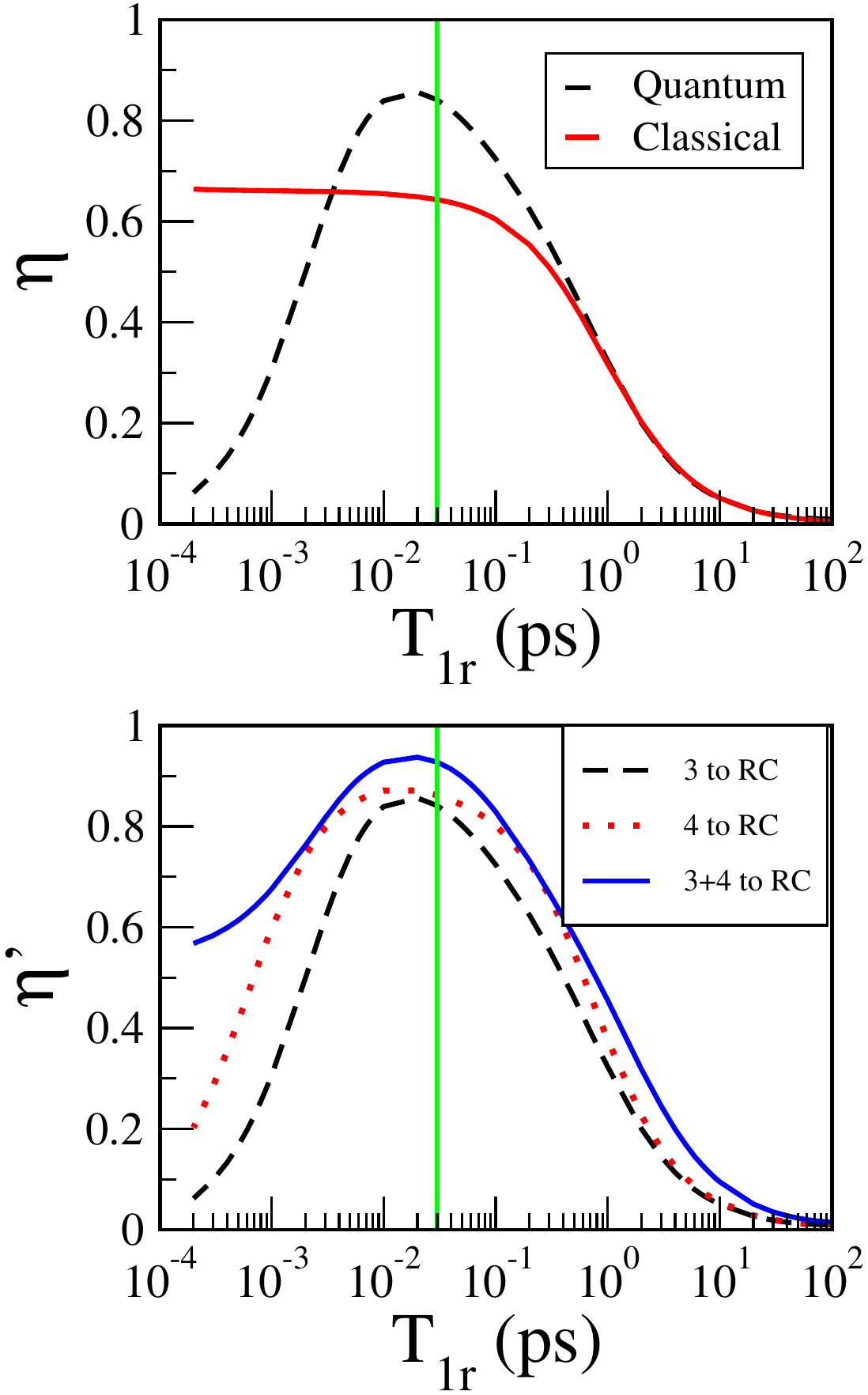}
\caption{ (Color online)
Upper panel :  quantum and classical efficiency computed 
at $t_{max} = 5  \ ps$,  
as a function of
$T_{1r}$ at room temperature dephasing rate, $\gamma_d(T= 300 \, K)$, see Eq.~(\ref{gammad}). 
Lower panel: efficiency computed at $t_{max} = 5 \ ps$,
using  Eq.~(\ref{eta2}),  
as a function of
 $T_{1r}$ , at room temperature dephasing rate,  $\gamma_d(T= 300 \, K)$,
for different coupling to RC.
The vertical green line  represents ST.
As initial conditions we choose 
$\rho(0)= (1/2) (|1 \rangle \langle 1| + |6 \rangle \langle 6|)$.
}
\label{fig3}
\end{figure}

The comparison between  classical and quantum behavior is shown in
Fig.~\ref{fig3} (upper panel). 
The classical dynamics leads to a very different dependence
of the efficiency on $T_{1r}$. Namely, the efficiency in 
the classical case  does not exhibit a maximum but 
simply decays with $T_{1r}$. This shows that the ST 
effect is due to quantum coherence
only. \\

\section{ Different coupling schemes}  
So far, we have considered
the site $3$ to be the only one coupled to the RC. However, it is not known for 
sure which sites are  connected
 to the RC, even though sites $3$
and $4$ are the
most likely candidates, since they are closest to the RC \cite{lloyd}.
As mentioned above, 
the non-Hermitian Hamiltonian formalism easily allows one
to describe different coupling schemes, which can be included
 in the effective 
Hamiltonian (\ref{Heff}) by properly choosing the coupling 
transition amplitude, $A_i^{^{RC}}$, between the
sites of the FMO complex and the RC. Note that while in the 
previous Section we indicated the channel in the RC with the number
3,  here we label it as  RC. 

\subsection{Coupling from site 3 and 4}
Since, to the best of our knowledge, it is not exactly known from
 experimental data how the energy transport occurs,
in the following we choose three different situations, showing that
the essential features of the phenomenon indicated in the previous
Section does not change too much. Specifically we consider:
\begin{itemize}
\item only site 3 is coupled to RC, so that we 
set $A_3^{^{RC}}=\sqrt{\hbar/2T_{1r}}$ 
(as done above), 
\item only site 4 is coupled to the RC, so we 
set  $A_4^{^{RC}}=\sqrt{\hbar/2T_{1r}}$, 
\item both sites $3$ and 
4 are coupled to the RC, so we set $A_3^{^{RC}}=A_4^{^{RC}}
=\sqrt{\hbar/2T_{1r}}$ . 
\end{itemize}
In a general setting the probability for the excitation 
to be in the RC at time $t_{max}= 5  \ ps$ cannot be computed using Eq.~(\ref{eta}), since by merely summing that expression for each site connected to the RC, we neglect interference effects. The efficiency should be computed using
\begin{equation}
\eta' (t_{max})= 1- Tr(\rho(t_{max})) - 
\frac{1}{T_{1}}
\int_0^{t_{max}}  \, dt \,    
Tr(\rho(t)).
\label{eta2}
\end{equation}
Here, $1-Tr(\rho(t_{max}))$ is the probability that the excitation leaves the
system by the time $t_{max}$. The last term in Eq.~(\ref{eta2}) is the probability that the excitation has been lost by recombination during this time.
If there is just one site coupled to the RC, then  Eq.~(\ref{eta2})
reduces to Eq.~(\ref{eta}).\\
In Fig.~\ref{fig3} (lower panel) we show that the efficiency is sensitive to
different coupling schemes. 
In particular, we notice that coupling through site $4$ achieves a greater efficiency than coupling  through site $3$. If both sites are coupled to the RC, then the efficiency is further improved, and the decay for small coupling times is smaller than that for a single coupled site. \\

\subsection{Efficiency {\it vs} position of the RC}

In the following, we consider a more realistic coupling scheme to RC,
namely the case in which  all  sites of the FMO system 
have an  electric dipole coupling to the same channel in the RC.
This assumption can be justified since 
in the RC there are the same  bacteriochlorophyll (BChl) molecules 
which compose the
FMO system. So, 
 it is reasonable to assume that the excitation is transferred to the RC by the same mechanism
that operates between the BChl molecules in the FMO system. 

The electric dipole transition amplitude from site $i$ to the RC can be written as: 
\begin{equation}
V_i^{^{RC}}=\frac{C}{R_{_{i,RC}}^3}
 \left[ \vec {\mu}_i \cdot \vec{ \mu}_{_{RC}} - 
3  (\vec{\mu}_i \cdot \hat{ R}_{_{i,RC}}) 
(\vec{\mu}_{_{RC}} \cdot \hat{ R}_{_{i,RC}}) \right],
\label{eldip}
\end{equation}
where $R_{_{i,RC}}$ is the distance from site $i$ to the RC,
${\mu}_i$ is the dipole moment at the site $i$, and  ${\mu}_{_{RC}}$ 
is the dipole moment assigned  to the RC. 
We take the position of the  BChls and their dipole moments from \cite{mohseni}.
Here, we assume  that the coupling strength between the
sites and the RC is equal to that between
the sites, so that: $C|\mu|^2= 134000$ $cm^{-1} (\AA)^3$,
as in Ref.~\cite{mohseni}.

In order to determine the coupling amplitude from 
site $i$ to the continuum of states in
the RC we  evaluate the transfer rate, $\kappa_i$ 
from the Fermi-golden rule \cite{qwalk}:
\begin{equation}
\kappa_i=\frac{ 2 \pi \rho_{_{RC}} |V_i^{^{RC}}|^2  }{\hbar},
\label{kappa}
\end{equation}
where $\rho_{_{RC}}$ represents the density of states in the RC.
It is interesting to observe that
an 
expression for the transfer rate similar to
(\ref{kappa})
can be also  obtained without 
perturbation theory, see for instance Ref.~\cite{rotter3},
where the continuum was  modeled as a semi-infinite lead. 
We can now determine the non-Hermitian Hamiltonian, Eq.~(\ref{Heff}), 
setting: 
$$
A_i^{^{RC}}= \sqrt{2 \pi \rho_{_{RC}} } \ V_i^{^{RC}}.
$$
 Note that now the coupling between 
the FMO complex and the RC depends on the position of the RC.
In order to determine how the efficiency of energy transfer
 depends on the position of the RC
w.r.t. the FMO complex, we assume ${{\vec \mu}}_{_{RC}}= 
{\vec \mu}_3$ and
place the RC in the same $y$ and $z$ position of site
$3$, as given in  \cite{mohseni}.  
The transport efficiency has been studied  by varying the 
distance, $d$, from the site 
$3$ along the $x$ direction.  
In order to compute the transition amplitude, $A_{i}^{^{RC}}$,
we need the density of states of the RC which,
to the best of our knowledge is not known experimentally.
 For this reason
we consider  different 
densities of states, respectively larger, equal
or smaller than the density of states of
 the FMO system,
$\rho_{_{FMO}} \simeq 1/D$,
 $D$
being  the mean level spacing of the FMO complex discussed  in Sec. (III).
\begin{figure}[ht!]
\vspace{0cm}
\includegraphics[width=8cm,angle=0]{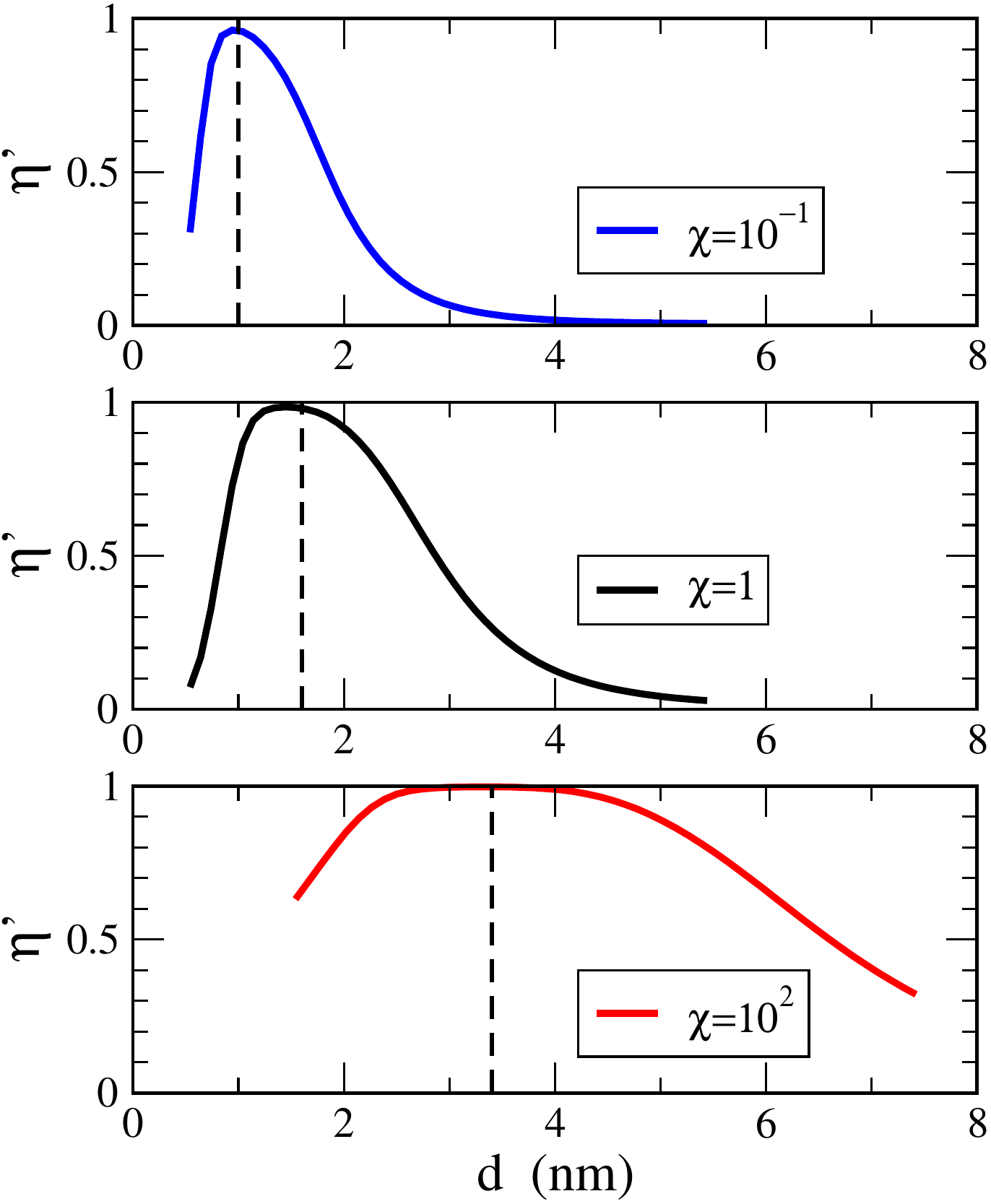}
\caption{ (Color online)
Efficiency, Eq.~(\ref{eta2}), computed at $t_{max} = 5 \ ps $ {\it vs} the
distance $d$  ($nm$) of the RC from site $3$ in the FMO complex
for different ratios $\chi = \rho_{_{RC}}/\rho_{_{FMO}}$. 
Here, an electric dipole coupling 
between the RC and the FMO system,
see Eq.~(\ref{eldip}), has been considered.
The vertical dashed lines,  obtained from Eq.~(\ref{R}), 
 represent the critical distances at which superradiance 
transition occurs.
As initial conditions we choose 
$\rho(0)= (1/2) (|1 \rangle \langle 1| + |6 \rangle \langle 6|)$.
Data in this figure refer to a room temperature dephasing rate, $\gamma_d(T= 300 \, K)$, 
see Eq.~(\ref{gammad}). 
}
\label{fig3b2}
\end{figure}
We  show in Fig.~(\ref{fig3b2}) 
how the efficiency, computed with Eq.~(\ref{eta2}) using Eq.~(\ref{HS}), 
varies as a function 
of the distance from the  RC to  site $3$ of the FMO system.
In Fig.~(\ref{fig3b2}) we consider  a 
room   temperature dephasing rate $\gamma_d(T = 300 \ K)$, for
different ratios, $\chi= \rho_{_{RC}}/ \rho_{_{FMO}} =0.1,1,10^2$,
respectively,  
from the upper to the lower panel.  As one
 can see, the optimal distance, which is the distance that
maximizes the efficiency, slowly depends on the density of states in the RC.

We can use the superradiant  criterium obtained in Eq.~(\ref{sr})
 to get an analytical
expression for the optimal distance of the RC from the FMO complex. 
Since site $3$ is the closest to the RC, we can use Eq.~(\ref{sr}) and 
we find  that the ST  occurs for $\hbar \kappa_3=D$.
Finally, from Eq.~(\ref{kappa}) we have:
\begin{equation}
d_{max} = (2 \pi  B^2)^{\frac{1}{6}} (\rho_{_{RC}} \rho_{_{FMO}})^{\frac{1}{6}},
\label{R}
\end{equation}
where,
\begin{equation}
B = C
 \left[ \vec {\mu}_3 \cdot \vec{ \mu}_{_{RC}} - 
3  (\vec{\mu}_3 \cdot \hat{ R}_{_{3,RC}}) 
(\vec{\mu}_{_{RC}} \cdot \hat{ R}_{_{3,RC}}) \right],
\label{eldip1}
\end{equation}
see Eq.~(\ref{eldip}).
Eq.~(\ref{R}) gives us the  distance at which superradiance transition occurs.
The critical distance obtained from Eq.~(\ref{R}) is shown 
in Fig.~(\ref{fig3b2})
as dashed vertical lines. As one can see, the estimate is very good. 
Note that changing the density of states in the RC from $\rho_{_{FMO}}/10$ to
$10^2  \rho_{_{FMO}}$ only  changes the optimal distance   
from $1 \ nm$ to $3 \ nm$. 
These distances are consistent  with available structural data for the 
RC-FMO complex, see for instance \cite{rc}. 
The result is remarkable since it  shows that the superradiant criterium  
 suffices to determine the optimal distance from the RC
to the FMO complex for a wide range of  values for the density
of states of the RC.

\section{Conclusion}

We have analyzed energy transport in the FMO system with the aid
 of a non-Hermitian Hamiltonian approach. 
This allows us  to take into account the
effect of the coupling of the FMO system to the reaction center
 in a consistent way,  not merely phenomenologically,
as is usually  done in the literature.
We have shown that by increasing the strength of the coupling to the
 reaction center, a superradiance transition occurs. 
This transition occurs at approximately
 the same  value of the coupling
 for which energy transport efficiency is maximal.
 Indeed, the superradiance transition  is due to coherent 
constructive interference between the paths
to the RC, and this effect enhances the rate of energy transfer.
Since the ST effect is due to quantum coherence, one might expect that
any consequences of ST would disappear in the presence of dephasing
and relaxation provided by the  thermal bath. 
On the contrary, we have shown that  the effect of superradiance 
survives in the presence of the thermal bath, and the
maximal efficiency only depends  weakly  on temperature.
We have also estimated the ST critical value analytically. 
 For coupling strengths of the FMO system to the RC near the critical
 one, where the 
superradiance transition takes place, 
 we obtained average energy transfer times comparable to experimental
 values (a few picoseconds).
 Finally, we took  into account a realistic coupling scheme between the FMO system and the RC, and 
derived from the superradiance condition  the 
analytical expression, Eq.~(\ref{R}), for the 
 optimal distance from the RC to the FMO complex. 
This analytical expression 
 depends on the density of states in the RC.
 Within a wide range of density of states
 the optimal distance which we obtained analytically is approximately a
 few nanometers and  is 
 consistent with available structural data on RC. 
  Note also that   Eq.~(\ref{R}) is valid for a generic Donor-Acceptor complex.

 Our analysis shows that  the
superradiance mechanism might play an important role in explaining
the efficiency of quantum transport in photosynthetic 
light-harvesting systems and in engineering artificial light harvesting systems.

{ \it Acknowledgments.}
This work has been supported by Regione Lombardia and CILEA Consortium
through a LISA 
(Laboratory for Interdisciplinary Advanced Simulation) Initiative
(2010/11)  grant [link:http://lisa.cilea.it].
 Support from grant D.2.2 (2010)
 from Universit\`a Cattolica
 is also acknowledged. The work by GPB was carried out under
the auspices of the National Nuclear Security Administration of the U.S. Department of Energy at the Los Alamos National Laboratory under Contract
No. DE-AC52- 06NA25396. MM has been supported by the NSERC Discovery Grant No. 205247. 
F.B., M.M. and G.P.B  thank   the Institut Henri Poincare (IHP) for partial support at the final stage of this work.


\end{document}